\documentclass[conference]{IEEEtran}
\IEEEoverridecommandlockouts
\usepackage{cite}
\usepackage{amsmath,amssymb,amsfonts}
\usepackage{graphicx}
\usepackage{textcomp}
\usepackage{xcolor}
\usepackage{algpseudocode}
\usepackage{mathtools} 
\usepackage{enumerate}
\usepackage[ruled, vlined, linesnumbered]{algorithm2e}

\def\BibTeX{{\rm B\kern-.05em{\sc i\kern-.025em b}\kern-.08em
    T\kern-.1667em\lower.7ex\hbox{E}\kern-.125emX}}
\begin{document}

\title{\huge Enhancing CoMP-RSMA Performance with Movable Antennas: A Meta-Learning Optimization Framework\\

}

\author{\IEEEauthorblockN{Ali Amhaz}

\and
\IEEEauthorblockN{ Shreya Khisa}

\and
\IEEEauthorblockN{Mohamed Elhattab}

\and
\IEEEauthorblockN{Chadi Assi}

\and
\IEEEauthorblockN{Sanaa Sharafeddine}
}

\maketitle
\begin{abstract} 
 This study investigates a downlink rate-splitting multiple access (RSMA) scenario in which multiple base stations (BSs), employing a coordinated multi-point (CoMP) transmission scheme, serve users equipped with movable antenna (MA) technology. Unlike traditional fixed-position antennas (FPAs), which are subject to random variations in wireless channels, MAs can be strategically repositioned to locations with more favorable channel conditions, thereby achieving enhanced spatial diversity gains.To leverage these advantages and maximize the achievable sum rate, we formulate an optimization problem that jointly determines the optimal transmit beamforming vectors at the BSs, the common stream allocation for different users, and the optimal positioning of the MAs, all while ensuring compliance with quality of service (QoS) constraints. However, the formulated problem is non-convex and computationally challenging due to the strong interdependence among the optimization variables. Traditional methods for solving large-scale optimization problems typically incur prohibitively high computational complexity. To address the above challenge, we propose a gradient-based meta-learning (GML) algorithm that operates without pre-training and is well-suited for handling large-scale optimization tasks. Numerical results demonstrate the effectiveness and accuracy of the proposed approach, achieving near-optimal performance (exceeding 97\% compared to the optimal solution). Moreover, the MA-enabled CoMP-RSMA model significantly outperforms conventional benchmark schemes, yielding performance gains of up to 190\% over the spatial division multiple access (SDMA) scheme and 80\% over the RSMA FPA-based model. Finally, the proposed approach is shown to mitigate the sum-rate limitations imposed by interference in SDMA, achieving superior performance with fewer BSs.
\end{abstract}


\section{Introduction}
\subsection{Background}
The evolution of wireless networks, particularly the sixth-generation (6G), will face an extraordinary surge in user density due to the explosive growth of connected devices and increasing wireless traffic. This trend coincides with the development of the Internet of Everything (IoE), which aims to connect billions of users, highlighting the pressing need for higher data rates, universal connectivity, and ultra-low latency \cite{wang2023road}. To address these challenges, researchers started studying cutting-edge multiple-access techniques and innovative network architectures. Multiple-input multiple-output (MIMO), a fundamental technology in 5G and next-generation wireless networks, provides an effective approach to enhance the degrees of freedom (DoFs) in communication systems. By leveraging multiple antennas at both the transmitter and receiver, MIMO significantly improves spectral efficiency, data rates, and signal reliability \cite{he2021cell}. This technology exploits spatial diversity and multiplexing gains, reducing interference and enhancing coverage. Nevertheless, traditional MIMO systems suffer from degraded performance due to inter-cell interference (ICI), which arises when signals from neighboring cells overlap. This phenomenon is more prominent for the cell edge users, that are more exposed to interference from adjacent cells \cite{8097026}.

In light of this, Coordinated Multi-Point (CoMP) transmission represents a suitable solution for mitigating ICI and improving spectral efficiency in multi-cell cellular networks. By leveraging the coherent transmission of multiple base stations (BSs), CoMP enables a form of spatial diversity that enhances the signal quality at user terminals. This is achieved through joint signal processing and cooperative transmission among geographically separated BSs \cite{10032129}. It leverages favorable propagation and channel hardening to enable efficient user multiplexing while minimizing the ICI \cite{8756668}. Nonetheless, the spectral efficiency at the user’s side remains fundamentally constrained by an upper bound, primarily due to the interference even with high transmission power. This interference arises from signals intended for different users, which cannot be eliminated, especially in scenarios involving space division multiple access (SDMA), a multiple access technique that exploits spatial separation of signals to serve multiple users simultaneously using the same frequency resources. In addition to the previous problem, striking the right balance between performance and complexity is essential because practical feasibility necessitates acquiring decent performance while reducing the number of active BSs participating in coordination \cite{8097026}.

Recently, movable antennas (MAs), an emerging technology, have started attracting attention due to their promising ability to enhance the DoF in the spatial domain. Unlike traditional fixed-position antennas (FPA)s, this advantage is harnessed through strategically positioning the antenna to enhance the quality of the channels, providing a better approach than antenna selection (AS), which mandates a large number of antennas and imposes extra implementation expenses \cite{zhu2023movable}. In addition, MA can continuously move, taking full advantage of spatial channel variations in all directions. In contrast, FPA systems, whether utilizing AS or not, are restricted to discrete antenna positions. With their maneuverability, MAs can hence provide great potential in suppressing interference, battling severe fading, and improving spatial diversity in CoMP scenarios owing to the improved beamforming design that can be obtained due to the ability to control the quality of wireless channels. 

Another important factor in improving spectral efficiency owes to the prudent selection of multiple access techniques. Non-traditional access techniques, specifically non-orthogonal multiple access (NOMA) and rate-splitting multiple access (RSMA), have received significant attention due to the ability to exploit the same frequency-time resources in the transmission process to numerous users \cite{10437094}, yielding enhanced spectral and energy efficiency from one side, and improved connectivity from the other side. Specifically, RSMA is a powerful multiple-access technique that allows multiple users to share the same resource block efficiently. By introducing a common stream that is decoded by multiple users, RSMA can enhance the spectral efficiency and mitigate interference. It leverages the successive interference cancellation (SIC) technique, enabling users to first decode and subtract the common stream before decoding their intended private messages. 

Given the advantages of these technologies and the associated challenges in resource management optimization, numerous studies have explored their feasibility by proposing system models that integrate MAs, CoMP, and RSMA across various scenarios, examining their interdependencies. The next subsection will provide an in-depth analysis of the latest state-of-the-art advancements in these fields.

\subsection{Motivation and State of the Art}

\par \textit{1) MA-enabled multiple-input single-output (MISO)/MIMO}: 
The concept and innovative architecture of MA were initially introduced in \cite{10318061}, where the authors conducted an in-depth analysis of MA. Their work presented a thorough modeling and performance evaluation of MA, benchmarking its effectiveness against FPA. The performance comparison results showed promising performance of MA over FPA in the situation where the number of channel paths increases due to more pronounced small-scale fading effects in the spatial domain. Subsequently, encouraged by the favorable results, several research studies have been performed utilizing MA in different scenarios and system models. The authors in \cite{10354003} presented an MA-enhanced multiple access channel (MAC) for uplink transmission where multiple users equipped with a single MA transmit their signals to the BS. Specifically, they formulated an optimization problem to minimize the total transmit power of the users by optimizing the MA positions and receiving beamforming vectors in the BS. This work is one of the initial investigations of MA-enhanced wireless systems. \par Additionally, the work in \cite{10243545} investigated the capacity of the MA-enabled MIMO communication system with the objective of maximizing the sum rate by jointly optimizing transmit and receive MA antennas. Besides, \cite{10851455} studied an MA-aided multi-user hybrid beamforming scheme with a sub-connected structure, where multiple movable sub-arrays can independently change their positions within different local regions.  With the objective of maximizing the sum rate, they jointly optimize digital beamforming, analog beamforming, and positions of sub-array. By invoking Lagrange multipliers, the penalty method, gradient descent, and alternating optimization (AO) are used to solve the challenging non-convex problem.  Meanwhile, the authors in \cite{10508218} suggested a graph-based approach to find the optimal MA positions compared to conventional optimization approaches. They proposed to sample the transmit region into discrete points whereas the continuous antenna position optimization problem is transformed to a discrete sampling point selection problem based on the point-wise channel information.
\par \textit{2) NOMA/RSMA-based MA-enabled Cellular Networks}: With the goal of improving connectivity and spectral efficiency in future wireless networks, several studies have started focusing on empowering MA-enabled networks with non-orthogonal multiple access techniques, specifically NOMA and RSMA. In \cite{10638767}, the authors examined deploying NOMA technology as an access technique for a system model comprising a set of users equipped with MAs. They formulated an optimization problem with the objective of maximizing the channel capacity by jointly determining the power allocation and the positions of the MAs. At the end, an AO approach was utilized to handle the non-convexity in the optimization problem. With a similar objective, the authors in \cite{10535440} studied an uplink NOMA scenario for a set of MA-enabled users transmitting to the BS. They jointly optimized the power control, decoding order, and the MAs' positions. Besides the work in \cite{gao2024movable} reported an MA-enabled wireless-powered communication network (WPCN) with NOMA where MA was flexibly utilized to harvest energy from downlink transmission of BS and to transfer information through uplink transmission to BS. An AO-based approach was utilized to solve the formulated non-convex problem to achieve a sub-optimal solution. 
\par Another promising non-orthogonal multiple access technique is the RSMA technology due to its ability to utilize the common stream in interference management and grant more flexibility. In this regard, the authors in \cite{zhang2024sumratemaximizationmovable} examined an MA-aided system in downlink RSMA where the BS was equipped with a set of linear MAs. The objective was to maximize the sum rate of the users by jointly optimizing the beamforming vectors, MAs positions, and the common stream portions. In order to handle the non-convexity in the presented optimization problem, they utilized a coarse-to-fine-grained searching (CFGS) algorithm. In \cite{10834523}, they studied a short packet transmission scheme empowered by RSMA and exploiting MA technology under Ultra-Reliable and Low-Latency Communication (URLLC) constraints. They formulated an optimization problem to maximize the sum rate by determining the optimal antenna positions and beamforming vectors. Moreover, in order to tackle the non-convexity in the formulated problem, an AO algorithm that was based on the Successive Convex Approximation (SCA) technique was utilized. 
\par \textit{3) Orthogonal multiple access (OMA)-based CoMP systems}:
Unlike traditional cellular networks, CoMP-based networks offer enhanced connectivity and reduced interference. In 3GPP LTE-Advanced, a work item on CoMP transmission and reception was launched in September 2011, becoming a key feature of Release 11 \cite{lee2012coordinated}. Exploiting the ability to coordinate between several BSs or distributed access points, they can offer higher spectral efficiency, better coverage, and improved reliability. 
\par The authors in \cite{10198677} studied a downlink CoMP transmission over LTE technology where a set of BSs are concurrently assisting a set of users. With the objective of maximizing the quality of experience (QoE) of the users, they proposed three novel deep-reinforcement learning approaches to determine the BS-user assignments while being able to adapt to the dynamic environment. Meanwhile, the authors in \cite{guidolin2014distributed} presented a distributed clustering algorithm that adapts the cluster configuration according to the user distribution and the average cluster size in LTE-based CoMP networks. Another work in \cite{mosleh2016proportional}, examined a downlink CoMP system by aligning transmit precoding with resource allocation with proportional-fair scheduling to achieve a balanced trade-off between spectral efficiency for both cell-edge and cell-average users. To solve the formulated optimization problem, a parallel SCA-based algorithm is introduced. 
\par To reduce the signalling overhead in backhaul, the authors in \cite{zhao2013coordinated} investigated a resource allocation problem by jointly designing transmit beamformers and user data allocation at BSs to minimize the backhaul user data transfer subject to given QoS and per-BS power constraints. In \cite{8097026}, the researchers examine a downlink cell-free massive MIMO scenario where a large number of access points serves a set of single-antenna users. They formulated an optimization problem to maximize the total energy efficiency under spectral and transmit power constraints by determining the optimal power allocation. Then, a set of mathematical approximations and an SCA algorithm were proposed to handle the non-convexity of the proposed optimization problem. 
\par \textit{4) NOMA/RSMA-based CoMP systems}:
In \cite{9737471}, the authors considered a joint transmission (JT) CoMP system model to assist a set of users that exist at the edge of the cells while enabling the cooperative transmissions between the users themselves to enhance the performance of the far users further. With the objective of maximizing the network sum rate, an optimization problem was formulated to jointly determine the power control scheme and the user's clustering. Moreover, in order to tackle the non-convex nature of the problem, they decomposed it into two problems where a one-to-one three-sided matching game approach was adopted to tackle the clustering problem. In \cite {muhammed2021resource}, the authors studied the user scheduling and power allocation problem to maximize the energy efficiency (EE) for NOMA in downlink CoMP networks while considering imperfect channel state information (CSI), imperfect SIC, and outages. In \cite{ali2018coordinated}, they proposed a generalized CoMP (GCoMP)-enabled NOMA scheme
that allows cooperation among distributed BSs to serve all users within the network coverage area. All users associated with a BS and using a particular frequency band form a single NOMA cluster. At the end,  they derived a closed-form expression for the probability of outage for a user with different orders of BS cooperation.
\par Another study \cite{8756668} considered the downlink RSMA-empowered CoMP scenario with the objective of maximizing the weighted sum rate (WSR) of the users. Accordingly, they formulated an optimization problem to determine the optimal beamforming design at the different BSs while respecting the QoS constraints. At the end, the problem was solved by utilizing an AO approach after transforming it to a Weighted Minimum Mean Square Error (WMMSE). Meanwhile, authors in \cite{elhattab2024coordinated}, presented a work for cooperative relaying in a downlink RSMA framework in a multi-cell CoMP scenario. They provided an investigation where cell-edge users were supported by cell-center users through cooperative relaying of the common stream to improve the signal quality of cell-edge users and mitigate the inter-cell interference through JT CoMP.

To the best of our knowledge, all the previously mentioned studies (along with the references cited therein) have primarily focused on one of two main approaches. The first category considers MA-enabled schemes within a single-cell scenario, overlooking the complexities introduced by multiple BSs. In such cases, inter-cell interference, which arises from overlapping coverage areas and uncoordinated transmissions, is not accounted for, potentially leading to significant performance degradation in practical multi-cell networks. The work in \cite{hu2024movable} represented the first trial of studying MAs with CoMP technology. The authors examined a system model that comprises an MA array for transmission along with a CoMP reception scheme to maximize the signal-to-interference-plus-noise ratio (SINR) received effectively. Nonetheless, the work focuses on the coordinated reception for a set of destinations equipped with a single FPA from one transmission BS ignoring the challenges induced by having multiple transmission sources. The second category focuses on CoMP models that rely on conventional FPA systems. While CoMP enhances spectral efficiency by enabling joint transmission and reception across multiple BSs, it remains fundamentally constrained by an upper bound on spectral efficiency. This limitation persists even under high transmission power, primarily due to other users' interference, which cannot be fully eliminated in traditional CoMP frameworks. Moreover, when several distributed BSs collaboratively serve users, maintaining an optimal balance between performance and complexity becomes crucial. To ensure practical feasibility, it is important to achieve high system performance while minimizing the number of active BSs involved in coordination. This is particularly essential for reducing system overhead, power consumption, and computational complexity, making efficient BS selection strategies a critical aspect of CoMP deployment.

In light of this and to the best of our knowledge, no previous study has tackled the RSMA-CoMP model empowered by MA technology on the user's side. The presented model takes advantage of the MA technology to overcome the challenges imposed by the upper bound limit on the spectral efficiency even with high transmit power due to the effect of interference by controlling the quality of channels (suppressing interference and combating fading effects). In addition, RSMA presented a great combination with MA technology, allowing to reach a higher achievable rate due to the ability to utilize the common stream and SIC process in interference management, offering more DoFs. In the end, this leads to fewer BSs needed to serve the user, decreasing complexity and saving resources. 

Nevertheless, implementing the aforementioned model is faced with challenges, as the MAs' positions need to be optimized to maximize the benefit from the transmission of BSs along with the beamforming vectors, and the common stream portions. Accordingly, the high coupling among the variables leads to a non-convex optimization problem that is hard to tackle using traditional optimization approaches. Hence, we adopt a gradient-based meta-learning (GML) approach based on three sub-neural networks (NNs) to optimize the variables iteratively.

\subsection{Contribution}
In this regard, the key contributions of this paper can be summarized as follows:
\begin{itemize}
  \item In this work and different from the existing literature, we study a novel model that embraces CoMP with MA technology as a way to overcome the fundamental limit of constraining the spectral efficiency to an upper bound by the ability to adjust the channel quality through interference suppression and battling fading. In addition, the RSMA technique was utilized to improve further the advantage of the MAs. The system model comprises a number of BSs that are intended to serve a set of users that are equipped with MAs in the downlink scenario. With the objective of maximizing the achievable sum rate, we formulate an optimization problem to determine the optimal beamforming design at the BSs, the common stream portions, and the MA positions while respecting the QoS requirements, the mobility region of the  MAs, and the power budget of all BSs. 
  \item The formulated optimization problem is non-convex in nature and hard to tackle using traditional optimization solvers because of the high coupling among the different optimization variables. To tackle this challenge, we propose a GML algorithm that functions without prior training and can efficiently manage large-scale optimization problems. 

  \item To optimize the three variables, namely, the beamforming design at the BSs, the common stream portions, and the MA positions, in the model, the meta-learning approach was designed to consist of three different sub-NNs to handle the three optimization variables in our scheme, namely, the beamforming design at the BSs, the common stream portions, and the MA positions. 
  \end{itemize}

To evaluate the performance of the model, we performed a set of extensive simulations. First, the convergence of the presented algorithm is illustrated, proving its ability to maximize our objective that deals with the achievable sum rate while respecting the different constraints. Then, the GML-based result is compared with the optimal solution, proving its accuracy and effectiveness in achieving near-optimal results. Next, the suggested model along with a set of benchmarks was investigated by varying the different networks' parameters, including the BSs' power budget, the minimum QoS requirements, the number of antennas at the BSs, the number of users, and the count of the utilized BSs in the cooperation process. The results revealed the advantage of MAs in enhancing the quality of channels by suppressing interference and battling fading, thus yielding higher spectral efficiency and overcoming the problem of the upper-bound limit explained above. Moreover, the advantages of MAs showed to be significantly amplified when accompanied by the RSMA scheme compared to the traditional SDMA model. In addition, the results demonstrated the ability of the presented model to achieve a similar sum rate to other benchmarks with fewer BSs saving the system resources.

\subsection{Outline}
This paper is organized into several sections. Section II introduces the system model and provides the rate analysis along with the adopted channel model for the MA technology. In Section III, we define the optimization problem. Section IV outlines the proposed GML algorithm, illustrating its components. Next, Section V presents the simulation results and performance validation. Finally, we present the conclusion in Section VI.
\section{System Model}

\begin{figure*}[!t]
    \centering    
\includegraphics[width=2.05\columnwidth]{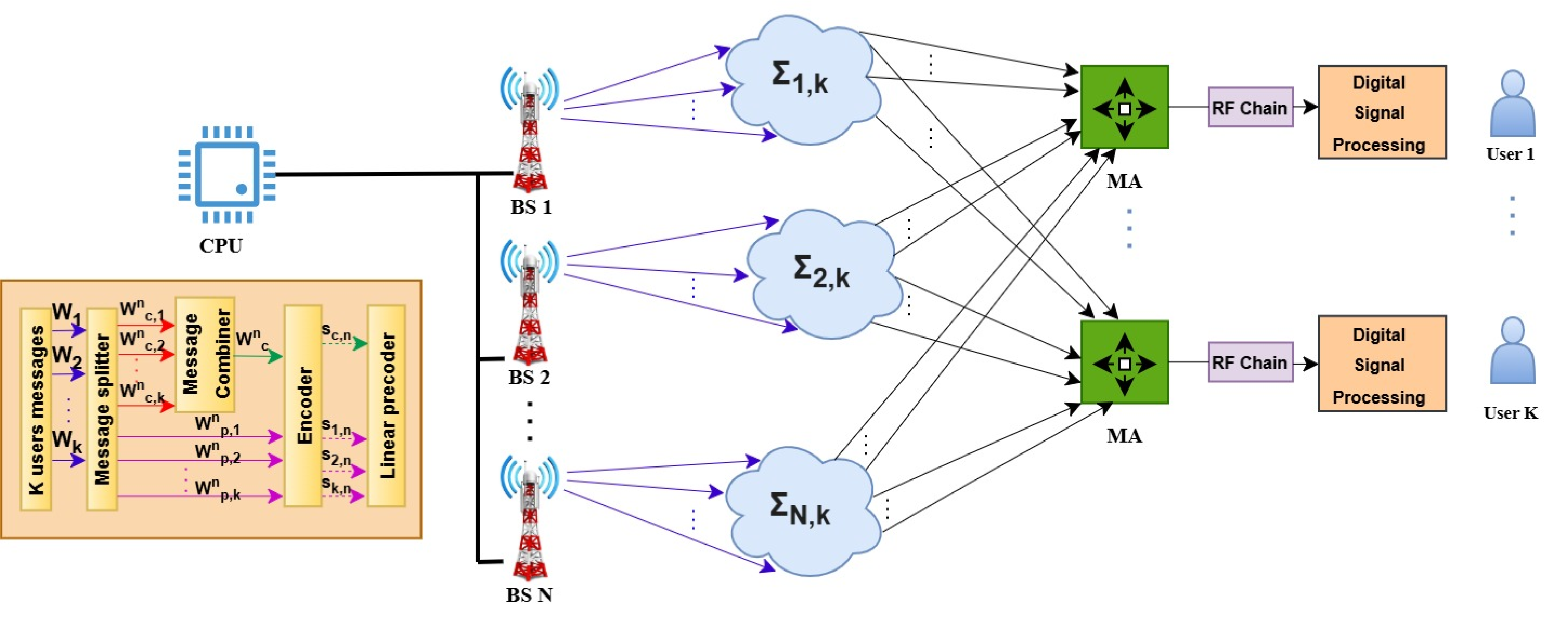}
    \caption{System model.}
    \label{fig_11}
\end{figure*}
\subsection{Network Model}

As illustrated in Fig.~\ref{fig_11}, we consider a downlink system consisting of $N$ BSs, each equipped with $I$ FPAs. The BSs are indexed by the set $\mathcal{N} = \{1, \dots, N\}$. Additionally, the system includes a set of $K$ users, denoted as $\mathcal{K} = \{1, \dots, K\}$, each equipped with a single MA. All BS antennas are connected to a remote server (RS)-based central processing unit (CPU), which can be deployed in a cloud environment. This will grant the ability to coordinate the transmission of the BSs. Hence, the different BSs will coherently serve the users over the same resources, empowering the CoMP system. We adopt the RSMA technology, in which each BS splits the messages of the users into common and private messages. The common messages of the different users will be combined to form a combined common message. Next, the combined common message will be encoded, creating a common stream. On the other hand, the private messages of the users will undergo a separate encoding process, forming a private stream that is unique for each user in the network. At the end, each BS will superimpose the common and private streams and transmit them to the different users. 
\subsection{Channel Model}
In this work, we adopt the field response channel model described in \cite{10318061}, \cite{10638767}. Accordingly, the channel response is determined by the combined contributions of coefficients from multiple transmission paths between the transceivers. Furthermore, the far-field channel condition is ensured between the BS and each user, as the signal propagation distance significantly exceeds the size of the MAs' moving region, leading at the end to an unchanging angle of arrivals (AoAs) and amplitudes of the complex path coefficients for multiple channel paths with the MA's displacement. Therefore, the only component affected by the mobility of the MAs is the phase of the multichannel paths. 

The channel between the BS $n$ and the user $k$ is denoted as $\boldsymbol{h}_{n,k}$. We denote by $O_{n,i} = [o_i, o_i]^T$ the origin of each antenna $i$ at the BS $n$. For any user $k$, the origin of the receiving antenna is denoted by $O_k = [0, 0]^T$. We express the number of transmit paths between the antenna $i$ of the BS $n$ and the user $k$ as $L_{i,n}^k$. On the flip side, the number of the received paths at the MA of the users is denoted as $Q_{i,n}^k$. The path response matrix (PRM) between the BS $n$ and the user $k$ is defined as $\boldsymbol{\Sigma}_{n,k}$, where $\boldsymbol{\Sigma}_{n,k}[a, b]$ corresponds to the channel response coefficient (CRC) between the $b$-th receive path and $a$-th transmit path of BS $n$ with the user $k$.

Now, for any user $k$, the difference in the phase of the propagating signal between the $m$-th transmit path of the antenna $i$ at BS $n$ (with position $t_{n,i} = [x_{n,i}, y_{n,i}]^T$) and the origin can be expressed as:  
\begin{equation}
    \rho^k_{m,n} = x_{n, i}^t \cos \theta^t_{m,k} \sin \phi^t_{m, k} + y_{n, i} \sin \theta_{m,k}^t
\end{equation}
where $\theta^t_{m,k}$ and $\phi^t_{m, k}$ denote the elevation and the azimuth of the angle of departure (AoD). Accordingly, and following \cite{10243545}, the transmit field response vector (FRV) of the $i$-th antenna of the BS $n$ can be expressed as follows:
\begin{align}
    &\textbf{g}_{n, k}(t_{n,i}) = \nonumber \\
    &\left[e^{j \frac{2\pi}{\lambda} \rho^t_{1, k} (t_{n,i})}, e^{j \frac{2\pi}{\lambda} \rho^t_{2, k} (t_{n,i})}, ..., e^{j \frac{2\pi}{\lambda} \rho^t_{L_{i,n}^k, k} (t_{n,i})} \right].
\end{align}
On the other hand, the receiving FRV at the user $k$ can be expressed as:
\begin{align}
    &\textbf{f}_k(r_k) = \left[e^{j \frac{2\pi}{\lambda} \rho^r_{1, k} (r_k)}, e^{j \frac{2\pi}{\lambda} \rho^r_{2, k} (r_k)}, ..., e^{j \frac{2\pi}{\lambda} \rho^r_{Q_{i,n}^k, k} (r_k)} \right], 
\end{align}
where $\rho^r_{d,k}$ denotes the difference in the phase of the propagating signal between the $d$-th receiving path of the MA of the user $k$ with location $r_k = [x_k, y_k]^T$ and the origin. It can be expressed as follows:
\begin{equation}
    \rho^r_{d,k} = x_k^r \cos \theta^r_{d,k} \sin \phi^r_{d, k} + y_k^r \sin \theta_{d,k}^r,
\end{equation}
where $\theta^r_{d,k}$ and $\phi_{d,k}^r$ denote the elevation and the azimuth of the AoA between the d-th path and the user $k$, respectively. Therefore, the channel $\boldsymbol{h}_k$ can be represented as follows:
\begin{align}
    &\boldsymbol{h}_{n,k} = \boldsymbol{f}_k^T (r_k). \boldsymbol{\Sigma}_{n,k} . \boldsymbol{G}_{n, k}, 
\end{align}
where the transmit field response matrix (FRM) $\boldsymbol{G}_{n, k}$ from the BS $n$ to the user $k$ is defined as  $\boldsymbol{G}_{n, k} = [\textbf{g}_{k, n}(t_{n, 1}), \textbf{g}_{k,n}(t_{n,2}), ..., \textbf{g}_{k, n}(t_{n, I}) ]$.


\subsection{Signal Model}
\noindent Utilizing the RSMA technology, each BS $n$ will divide the message of every user $k \in \mathcal{K}$ into a private message $W_{p,k}^n$ and a common message $W_{c,k}^n$. Next, each BS will combine the common messages, forming the message $W_{c}^n$, after which it goes through an encoding process yielding the stream $\boldsymbol{s}_{c,n}$. In parallel, each private message will experience a separate encoding process, forming the streams $\boldsymbol{s}_{k,n}, \forall k \in \mathcal{K}, \forall n \in \mathcal{N} $. Based on this, we define  $\boldsymbol{s}_n=[\boldsymbol{s}_{c, n},\boldsymbol{s}_{k, n}, \forall k \in \mathcal{K}]^T$ 
to be the total data stream that is being sent by the BS $n$. Finally, each BS will perform linear precoding for the data streams through the precoding vectors   $\boldsymbol{P}_n=[\boldsymbol{p}_{c,n},\boldsymbol{p}_{k, n},\forall k \in \mathcal{K} ]$ where $\boldsymbol{p}_{c,n}$ corresponds to the precoding vectors of the common stream at BS $n$ while $\boldsymbol{p}_{k,n}$ denotes the precoding vectors for the private stream of user $k$. Hence, the transmitted signal by the BS $n$ can be defined as:
\begin{equation}
\boldsymbol{x}_n = \boldsymbol{s}_{c,n}\boldsymbol{p}_{c,n} + \sum_{k=1}^K \boldsymbol{s}_{k,n}\boldsymbol{p}_{k,n}.  %
\end{equation}
\noindent The received signal by the user $k$  from the BSs in the direct transmission phase can be defined as: 
\begin{equation}  
y_{k}=\sum_{n=1}^{N} \boldsymbol{h}_{n,k}^H \boldsymbol{x}_n +n_{k}.
\end{equation}
\noindent The decoding process at user $k$ starts by decoding first the common stream. Hence, the SINR to decode the common stream at user $k$ could be expressed as:
\begin{equation}
    \Lambda_{c,k} = \frac{\left|\sum_{n = 1}^{N}\boldsymbol{h}_{n,k}^H \boldsymbol{p}_{c,n}\right|^2}{ \sum_{n=1}^N \sum_{k=1}^K\left|\boldsymbol{h}_{n,k}^H \boldsymbol{p}_{k,n}\right|^2  + \sigma^2}.
\end{equation}
Accordingly, the rate to decode the common message at user $k$ can be defined as:
\begin{equation}
    R_{c,k} = \log_2(1 +  \Lambda_{c,k}) .
    \label{common_r}
\end{equation}
After removing the common message using the SIC technique, the private message of user $k$ can be decoded with the SINR  expressed as:
\begin{equation}
    \Lambda_{p,k} = \frac{\left|\sum_{n = 1}^{N}\boldsymbol{h}_{n,k}^H \boldsymbol{p}_{k,n}\right|^2}{\sum_{n=1}^N \sum_{k' = 1, k'\neq k} ^K\left|\boldsymbol{h}_{n,k'}^H \boldsymbol{p}_{k',n}\right|^2 + \sigma^2},
\end{equation}
 The corresponding achievable rate can be defined as:
\begin{equation}
    R_{p,k} =  \log_2 ( 1 +  \Lambda_{p,k}).
    \label{private_r}
\end{equation}
\noindent In order to ensure successful decoding of the common stream among all the users, the achievable data rate of the  common stream should be constrained by the minimum of all the achievable common stream rates of all users, and it can be denoted as:
\begin{equation}
    R_c = \min(R_{c,1},.., R_{c,k}).
\end{equation}
\noindent Now, the rate $R_c$  will be divided among the users into the portions $c_{k}$, $\forall k \in \mathcal{K}$ such that:
\begin{equation}
    \sum_{k =1}^Kc_{k} \leq R_c.
\end{equation}
Finally, the total achievable rate at the user $k$ after adding the private and common rates can be expressed as follows:
\begin{align}
&R_{k} = R_{p,k} + c_{k}, ~~~ k \in \mathcal{K}.
\end{align}
\vspace{-0.4cm}
\section{Problem Formulation}
In this section, a mathematical problem is formulated with the objective of maximizing the achievable network sum rate by jointly optimizing the precoding matrix of the BSs to be the set of $\boldsymbol{P}_n$, $ \forall n \in \mathcal{N}$, i.e., $\boldsymbol{P} = \{\boldsymbol{P}_n|n\in \mathcal{N}\}$, the common stream portion vector $\boldsymbol{c}_k$ $, \forall k \in \mathcal{K}$, i.e., $\boldsymbol{C} = \{\boldsymbol{c}_k|k\in \mathcal{K}\}$, and the movable antenna positions for the different users $\boldsymbol{r}_k$ $ \forall k \in \mathcal{K}$, i.e., $\boldsymbol{L} = \{\boldsymbol{r}_k|k\in \mathcal{K}\}$. Hence, the optimization problem is expressed as
\begin{subequations}
\begin{align}
\mathcal{P}_1: \max_{\boldsymbol{P}, \boldsymbol{C}, \boldsymbol{L  }}&  \sum_{k=1}^K\left(c_k + R_{p,k}\right),   \ \;   \\
\textrm{s.t.}  \ \;  & \sum_k^K \left(\boldsymbol{p}_{k, n}^H \boldsymbol{p}_{k, n} +\boldsymbol{p}_{c,n}^H \boldsymbol{p}_{c_n}\right)\leq P_M^n ,\forall n \in \mathcal{N} \label{C_77}\\
   & \left(c_k + R_{p,k}\right) \ge R_{th}, \forall k \in \mathcal{K} \label{rth},\\
   & c_k \ge 0, \quad \forall k \in \mathcal{K} \label{e1},\\
   & \sum_{k=1}^K{c_k} \le R_c \label{e2},\\
   & \boldsymbol{r}_k \in \mathcal{D}_{r_k}, \forall k \in \mathcal{K}, \label{rx1}
\end{align} 
\label{optim}\end{subequations}where constraint \eqref{C_77} restricts the power budget at all the BSs, constraint \eqref{rth} corresponds to the QoS requirements for the different users with a rate threshold $R_{th}$, meanwhile \eqref{e1} and \eqref{e2} correspond to the common stream constraints described in Section II.C. Finally, constraint \eqref{rx1} restricts the maneuverability of the MA for all the users in which the mobility region of user $k$ is defined as $D_{r_k} = A \times A$ in a 2D space. In the next section, we go over our proposed solution approach. 
\section{Solution Approach}
The high coupling between the different decision variables in the optimization problems renders a non-convex problem that is difficult to tackle using traditional optimization solvers. In light of this, we adopt a GML method to solve large-scale optimization problems. Unlike the traditional deep-learning (DL)-based approach \cite{9805773}, where the different optimization variables and parameters are fed up as an input for the neural network to generate the optimal values as output, our approach depends on feeding the gradients $\Delta_{\boldsymbol{P}}R_{\boldsymbol{P}}$, $\Delta_{\boldsymbol{L}}R_{\boldsymbol{L}}$, and $\Delta_{\boldsymbol{C}}R_{\boldsymbol{C}}$ (based on the objective function $R_a$, $\forall a \in \{\boldsymbol{P}, \boldsymbol{L}, \boldsymbol{C}\}$) to the different neural networks in the model and outputting the gradients $\Delta{\boldsymbol{P}}$,$\Delta{\boldsymbol{L}}$, and $\Delta{\boldsymbol{C}}$ that are added over the initial values of the variables $\boldsymbol{P}$, $\boldsymbol{C}$, and $\boldsymbol{L}$. Such an approach will yield a better interpretable model compared to the traditional NN models, which are usually black boxes. In addition, the fact that the presented model depends on the gradients rather than the actual values allows the extraction of higher-order information for the variables $\boldsymbol{P}$, $\boldsymbol{C}$, and $\boldsymbol{L}$ with lower complexity \cite{10623434}. 


\subsection{Gradient-based Meta Learning (GML) Architecture}
Typical data-driven meta-learning approaches often require extensive offline pre-training followed by additional online adaptation and refinement, such as in model-agnostic meta-learning, leading to significant variations in the data distribution. This problem is accompanied by extra energy consumption due to the need to apply large-scale pretraining and adaptation processes, yielding serious troubles for dynamic scenarios and latency-critical applications. In light of these problems, we introduce a model-driven meta-learning framework that eliminates the need for pre-training and demonstrates strong robustness. It focuses its optimization primarily on the search trajectory rather than individual variables. 
Those two components can be characterized by three different iteration layers under which the optimization process takes place. They are defined as follows:

\subsubsection{Inner Iteration}
At this stage, the optimization of the different variables $\boldsymbol{P}$, $\boldsymbol{L}$, and $\boldsymbol{C}$ takes place cyclically. In this regard, each variable is assigned to a sub-network as follows:

\begin{itemize}
  \item \textbf{Precoding Network (PN)}: This network optimizes the precoding matrix at the different BSs.  
\item \textbf{Common Network (CN)}: This network optimizes the common stream portions $C$.  
\item \textbf{Movable Antennas Network (AN)}: This networks optimizes the positioning of the different MAs at the users.
  \end{itemize}

For all sub-networks, the variables are updated sequentially, and the optimization variable of a certain network is obtained from the initial points. On the other hand, the other two variables are extracted from the other networks. We can define the update process as follows:
\begin{align}
&\boldsymbol{C}^*=R(\boldsymbol{C}^{(i,j)},\boldsymbol{P}^*,\boldsymbol{L}^*), \\
   & \boldsymbol{P}^*=R(\boldsymbol{P}^{(i,j)},\boldsymbol{C}^*,\boldsymbol{L}^*),\\
   & \boldsymbol{L}^*=R(\boldsymbol{L}^{(i,j)} ,\boldsymbol{C}^*,\boldsymbol{P}^*).
\end{align}where $\boldsymbol{C}^{(i,j)}$, $\boldsymbol{P}^{(i,j)}$, and $\boldsymbol{L}^{(i,j)}$ correspond to the values of $\boldsymbol{C}$, $\boldsymbol{P}$, and $\boldsymbol{L}$ in the outer itration of $j$ and the inner iteration of $i$. We provide below the details on the different sub-networks.
\par \textit{\textbf{Precoding Network (PN)}}:
With the goal of optimizing the precoding vectors at all the BSs in the $j$-th outer iteration and $i$-th inner iteration, the objective function of maximizing the achievable sum rate at the communication users can be written as $R(\boldsymbol{P}^{(i,j)},\hat{\boldsymbol{C}},\hat{\boldsymbol{L}})$ where the terms $\hat{\boldsymbol{C}},\hat{\boldsymbol{L}}$ denote the initial or updated value for the common portion and MAs matrices, respectively. Initially, the achievable sum rate is calculated and accordingly, the gradient of $\boldsymbol{P}^{(i,j)}$ is obtained and is injected into the neural network with output $\Delta \boldsymbol{P}^{(i,j)}$ that will be added to $\boldsymbol{P}^{(i,j)}$ as follow:
\begin{equation}
\begin{split}
&\boldsymbol{P}^{(i+1,j)}=\boldsymbol{P}^{(i,j)}+\Delta\boldsymbol{P}^{(i,j)}. \label{pup}
\end{split}
\end{equation}
\par \textit{\textbf{Common Network (CN)}}:
Following the same strategy as the previous network, we denote by $R(\boldsymbol{C}^{(i,j)},\hat{\boldsymbol{P}}, \hat{\boldsymbol{L}})$ the objective at the $i$-th inner iteration and $j$-th outer iteration, where the terms $\hat{\boldsymbol{P}}$ denotes the initial or updated value for the precoding matrix of the BSs. Hence, after calculating the gradient for $\boldsymbol{C}$ with respect to the achievable sum rate, the output of the corresponding neural network $\Delta \boldsymbol{C}^{(i,j)}$ is added to $\boldsymbol{C}^{(i,j)}$ as follow:
\begin{equation}
\begin{split}
&\boldsymbol{C}^{(i+1,j)}=\boldsymbol{C}^{(i,j)}+\Delta\boldsymbol{C}^{(i,j)}. \label{cup}
\end{split}
\end{equation}
\par \textit{\textbf{Movable Antennas Network (AN)}}:
Similar to the networks $\boldsymbol{PN}$ and $\boldsymbol{CN}$, the objective at the iteration $i$ of the inner loop, and $j$ of the outer one can be expressed $R(\boldsymbol{L}^{(i,j)},\hat{\boldsymbol{C}},\hat{\boldsymbol{P}})$. In addition, after obtaining the gradient $\Delta \boldsymbol{L}^{(i,j)}$ from the neural network, the update process can be defined as:
\begin{equation}
    \begin{split}
&\boldsymbol{L}^{(i+1,j)}=\boldsymbol{L}^{(i,j)}+\Delta\boldsymbol{L}^{(i,j)}. \label{cup}
\end{split}
\end{equation}
\subsubsection{Outer Iteration}

During each outer iteration, $N_i$ inner iterations help in accumulating the meta-loss. This meta-loss function must integrate both the goal of maximizing the sum rate and satisfying the constraints required to maintain the feasibility of the formulated optimization problem \eqref{optim}. In light of this, the meta-loss function can be defined as follows:
\begin{equation}
    \mathcal{L}^j= \mathcal{L}_{rate}^j + \mathcal{L}_{threshold}^j + \mathcal{L}_{common1}^j + \mathcal{L}_{common2}^j + \mathcal{L}_{MA}^j 
    \label{loss}
\end{equation}
\noindent where the variable $\mathcal{L}_{rate}$ corresponds to the loss of the objective function which is expressed as the negative of the achievable sum rate as follows:
\begin{equation}
    \mathcal{L}_{rate}^j =  - \sum_{k=1}^{K} R_k.
\end{equation}
\noindent The variable $\mathcal{L}_{threshold}$ is added in order to guarantee satisfying the constraint \eqref{rth} and it is defined as:
\begin{equation}
    \mathcal{L}_{threshold}^j= \sum_{k=1}^K \zeta_1 \lambda (R_k), 
\end{equation}
\noindent where $\zeta_1$ is a regularization parameter and $\lambda(.)$ is an indicator function defined as:
\begin{equation}
    \lambda  = \begin{cases}0, & \mbox{if } \mbox{$R_{th}-R_k \leq 0$}, \\  1, & \mbox{} \mbox{otherwise}. \end{cases}
\end{equation}
\noindent To satisfy the constraint $\eqref{e1}$ and $\eqref{e2}$ that corresponds to the common stream, the terms $\mathcal{L}_{common1}$, $ \mathcal{L}_{common2}$ are defined as follow:
\begin{equation}
    \mathcal{L}_{common1}^j = \sum_{k=1}^{K} \zeta_2 I_1 (c_k),
\end{equation}
\begin{equation}
    \mathcal{L}_{common2}^j = \zeta_3 I_2 ( \sum_{k=1}^K c_k),
\end{equation}
\noindent where $\zeta_2$ and $\zeta_3$ are regularization parameters and $I_1(.)$ and $I_2(.)$ are indicator functions defined as:

\begin{equation}
    I_1  = \begin{cases}0, & \mbox{if } \mbox{$c_k \leq 0$}, \\  1, & \mbox{} \mbox{otherwise}, \end{cases}
\end{equation}
\begin{equation}
    I_2  = \begin{cases}0, & \mbox{if } \mbox{$\sum_{k=1}^K c_k \ge R_c$}, \\  1, & \mbox{} \mbox{otherwise}. \end{cases}
\end{equation}
Now, the term $\mathcal{L}_{MA}$ is included to restrict the mobility of the MAs in the designated area as illustrated in the constraint \eqref{rx1}. It is expressed as:
\begin{equation}
    \mathcal{L}_{MA}^j= \sum_{k=1}^K \zeta_4 I_3 (\boldsymbol{r}_k), 
\end{equation}
where $\zeta_4$ is a regularization parameter and $I_3(.)$ is an indicator function denoted as follow:
\begin{equation}
    I_3  = \begin{cases}0, & \mbox{if } \mbox{$\boldsymbol{r}_k \in \mathcal{D}_{r_k}$}, \\  1, & \mbox{} \mbox{otherwise}. \end{cases}
\end{equation}
Although the terms $\mathcal{L}_{threshold}, \mathcal{L}_{common1}, \mathcal{L}_{common2}, \mathcal{L}_{MA} $ that was introduced to the loss function $\mathcal{L}^j$ allows enforcing the constraints \eqref{rth}, \eqref{e1}, \eqref{e2}, and \eqref{rx1}, the constraint \eqref{C_77} which is responsible for respecting the power budget at the different BSs is still not tackled. In light of this and following \cite{loli2024metalearningbasedoptimizationlarge}, we utilize a normalization technique that is based on projecting the updated matrix $\boldsymbol{P}^{*}$ that consists of all the precoding matrices of the different BSs $\boldsymbol{P}^*_n$ $\forall n \in \mathcal{N}$. In this regard,  we apply the following normalization technique to all $\boldsymbol{P}^*_n$ as follows:

\begin{equation}
        U(\boldsymbol{P}_n^*)  = \begin{cases}\boldsymbol{P}_n^*, & \mbox{if } \mbox{$Tr(\boldsymbol{P}_n^*(\boldsymbol{P}_n^*)^H) \leq P_M^n$}, \\  \sqrt{\frac{P_M^n}{Tr(\boldsymbol{P}_n^*(\boldsymbol{P}_n^*)^H) }}\boldsymbol{P}^*_n, & \mbox{otherwise}. \end{cases}
        \label{normalized}
\end{equation}

\subsubsection{Epoch Iteration}
At this level, the update process of the parameter of the neural network takes place. It consists of $N_o$ outer iterations after which the different losses are summed and the average is obtained as below:
\begin{equation}
    \bar{\mathcal{L}}=\frac{1}{N_o}\sum_{j=1}^{N_o}\mathcal{L}^j.
\end{equation}
Next, backpropagation is performed, and the Adam optimizer is employed to update the neural networks of the presented model, as illustrated below:
\begin{equation}
\theta_{\textbf{C}}^*=\theta_{\textbf{C}}+\beta_{\textbf{C}}.\textrm{Adam}(\Delta_{\theta_{\textbf{C}}}\mathcal{\bar{L}},\theta_{\textbf{C}}),\label{w}
\end{equation}
\begin{equation}
\theta_{\textbf{P}}^*=\theta_{\textbf{P}}+\beta_{\textbf{P}}.\textrm{Adam}(\Delta_{\theta_{\textbf{P}}}\mathcal{\bar{L}},\theta_{\textbf{P}}),\label{p}
\end{equation}
\begin{equation}
\theta_{\boldsymbol{L}}^*=\theta_{\boldsymbol{L}}+\beta_{\boldsymbol{L}}.\textrm{Adam}(\Delta_{\theta_{\boldsymbol{L}}}\mathcal{\bar{L}},\theta_{\boldsymbol{L}}),\label{phi}
\end{equation}
where $\theta_{\textbf{C}}$, $\theta_{\textbf{P}}$, and $\theta_{\textbf{L}}$ represent the neural network parameters for the sub-networks \textbf{CN}, \textbf{CP}, and \textbf{AN}, respectively with the learning rates  expressed as $\beta_{\textbf{C}}$, $\beta_{\textbf{P}}$, and $\beta_{\boldsymbol{L}}$, respectively.  The pseudo-code of the presented steps are illustrated in \textbf{Algorithm} 1.

\begin{algorithm}[!t]
\footnotesize
\caption{Proposed algorithm}\label{alg:3}
 Randomly initialize $\theta_{\textbf{P}},\theta_{\textbf{C}},\theta_{\boldsymbol{L}},\textbf{P}^{(0,1)}$, $\textbf{C}^{(0,1)},\boldsymbol{L}^{(0,1)}$;\\
 Initialize inner iterations, $N_i$, outer iterations, $N_o$, and epoch iterations, $N_e$;\\
\For {$e={1,2,...,N_e}$}{
$\bar{\mathcal{L}}=0$;\\
MAX$=0$;\\
\For {$j=1,2,...N_o$}{
  $\textbf{C}^{(0,j)}=\textbf{C}^{(0,1)}$ ;\\
   $\textbf{P}^{(0,j)}=\textbf{P}^{(0,1)}$;\\
    $\boldsymbol{L}^{(0,j)}=\boldsymbol{L}^{(0,1)}$;\\
      \For {$i= 1,2,...,N_i$}{
      $R_{\textbf{C}}^{(i-1,j)}=R(\textbf{C}^{(i-1,j)},\textbf{P}^*,\boldsymbol{L}^*)$;\\
      $\Delta \textbf{C}^{(i-1,j)}$= $\textbf{\textrm{CN}}(\Delta_{\textbf{C}}R_{\textbf{C}}^{(i-1,j)})$;\\
      $\textbf{C}^{(i,j)}=\textbf{C}^{(i-1,j)}+\Delta \textbf{C}^{(i-1,j)}$;
      }
      $\textbf{C}^*=\textbf{C}^{(N_i,j)}$;\\
        \For {$i= 1,2,...,N_i$}{
      $R_{\textbf{P}}^{(i-1,j)}=R(\textbf{P}^{(i-1,j)},\textbf{C}^*,\boldsymbol{L}^*)$;\\
      $\Delta \textbf{P}^{(i-1,j)}$= $\textbf{\textrm{PN}}$
      $(\Delta_{\textbf{P}}R_{\textbf{P}}^{(i-1,j)})$;\\
      $\textbf{P}^{(i,j)}=\textbf{P}^{(i-1,j)}+\Delta \textbf{P}^{(i-1,j)}$;
      }
      $\textbf{P}^*=\textbf{P}^{(N_i,j)}$;\\
      Apply the normalization as in \eqref{normalized}\\
        \For {$i= 1,2,...,N_i$}{
      $R_{\boldsymbol{L}}^{(i-1,j)}=R(\boldsymbol{L}^{(i-1,j)},\textbf{C}^*,\textbf{P}^*)$;\\
      $\Delta \boldsymbol{L}^{(i-1,j)}$= $\textbf{\textrm{AN}}$
      $(\Delta_{\boldsymbol{L}}R_{\boldsymbol{L}}^{(i-1,j)})$;\\
      $\boldsymbol{L}^{(i,j)}=\boldsymbol{L}^{(i-1,j)}+\Delta \boldsymbol{L}^{(i-1,j)}$;
      }
    $\boldsymbol{L}^*=\boldsymbol{L}^{(N_i,j)}$;\\
   Calculate $\mathcal{L}^j$ as in \eqref{loss};\\
   $\bar{\mathcal{L}}=\bar{\mathcal{L}}+\mathcal{L}^j$;\\
   \textbf{If} {$-\mathcal{L}^j>MAX$}\\
        \quad MAX=$-\mathcal{L}^j$;\\
        \quad $\textbf{W}_{opt}=\textbf{W}^*$;\\
        \quad $\textbf{P}_{opt}=\textbf{P}^*$;\\
        \quad $\boldsymbol{C}_{opt}=\boldsymbol{C}^*$;\\
   \textbf{end if}
   }
   
$\bar{\mathcal{L}}=\frac{1}{N_o}*\bar{\mathcal{L}}$;\\
update $\theta_{\textbf{P}}$ , $\theta_{\textbf{C}}$, and $\theta_{\textbf{L}}$ , \\

}
return $\textbf{C}_{opt},\textbf{P}_{opt},\boldsymbol{L}_{opt}$
\end{algorithm}
\subsection{Computational Complexity Analysis}
The time complexity of the presented model can be illustrated as follows \cite{10623434}:

For the Precoding Network (\textbf{PN}), the complexity is primarily influenced by the computations required for evaluating the rate for all the users along with the neural networks. To obtain the rate, the equations \eqref{common_r} and \eqref{private_r} should be evaluated. Equation \eqref{common_r} involves multiplying the terms $\boldsymbol{h}_{n,k}$ with $\boldsymbol{p}_{c,n}$ and $\boldsymbol{p}_{k,n}$ leading to a computational cost of $O(I^2)$. This operation will be repeated $K$.$N$ times in the expressions $\sum_{n=1}^N \sum_{k=1}^K|\boldsymbol{h}_{n,k}^H \boldsymbol{p}_{k,n}|^2$ leading to a total  time complexity of $O(K.N.I^2)$. Similarly, the complexity for equation \eqref{private_r} can be obtained, leading to a complexity of $O((K-1).N.I^2)$. Thus, the final complexity of adding the two terms can be calculated as $O(K.N.I^2)$. Then, those two operations are done for K users, leading to a complexity of $O(K^2.N.I^2)$. 
In addition to the SINR expressions, normalizing the beamforming vectors has a computational cost that can be described as $O(I.K^2)$. Now, the complexity related to the NNs can be defined as $O(K^2)$ \cite{10623434}. Hence, combining all the previous complexities yields the expression  $O(K^2.N.I^2)$ + $O(I.K^2)$ + $O(K^2)$ = $O(K^2.N.I^2)$.

Similarly, the complexities associated with the Common Network (\textbf{CN}) and Movable Antennas Network (\textbf{AN}) can be obtained using a similar strategy, leading to the complexity for both components  $O(K^2.N.I^2)$. 
Finally, by taking into account the inner, outer, and epoch iterations, the final time complexity of the proposed meta-learning scheme can be defined as $N_e N_o N_i (O(K^2.N.I^2) + O(K^2.N.I^2) + O(K^2.N.I^2))$ =  $N_e N_o N_i (O(K^2.N.I^2))$.

\begin{figure}[!t]
    \centering    \includegraphics[width=1.1\columnwidth]{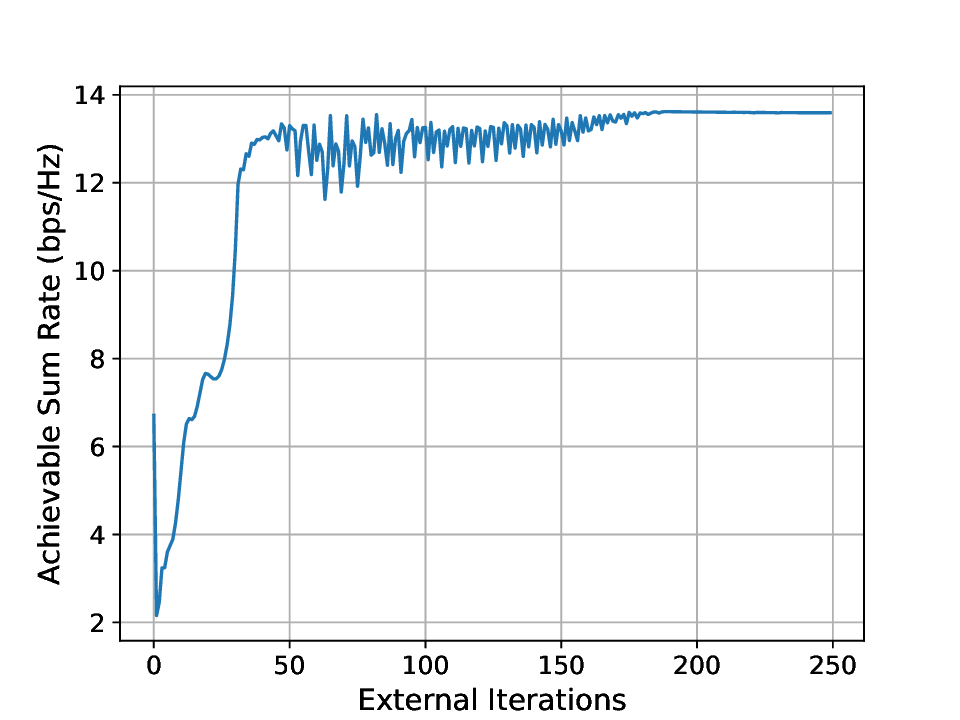}
    \caption{Convergence of the proposed algorithm.}
	\label{convv}
\end{figure}

\section{Numerical Evaluation}

\begin{table}[t]
\caption{Simulation Parameters}
\centering
\begin{tabular}{||l||c||l||c||}
    \hline
        Parameters & Values & Parameters & Values \\ \hline
        ~$P_M^n ,\forall n \in N$ & ~33dBm& ~$\lambda$ & ~0.01 \\ \hline
        ~$L_{i,n}^k$ $ \forall n \in \mathcal{N}, \forall k \in \mathcal{K}$& ~6 & ~$\sigma^2$ & -100 dBm/Hz \\ \hline
        ~$Q_{i,n}^k$ $ \forall n \in \mathcal{N}, \forall k \in \mathcal{K}$ & ~6 & ~$A$ & ~2$\lambda$ \\ \hline
        ~$N$ & ~2 & ~$I$ & ~4 \\ \hline
        ~$R_{th}$ & ~0.6bps/Hz & ~$\alpha$ & ~3.9\\ \hline
        ~$K$ & 4 & ~$\beta_{\textbf{C}}$ & ~$1e^{-3}$\\ \hline
        ~$\beta_{\textbf{P}}$ & ~$1.6e^{-3}$ & ~$\beta_{\boldsymbol{L}}$ & ~$1e^{-5}$\\ \hline
        ~$\zeta_1$ & ~$0.0001$ & ~$\zeta_2$ & ~$0.01$\\ \hline
        ~$\zeta_3$ & ~$1.001$ & ~$\zeta_4$ & ~$0.0001$\\ \hline
  \end{tabular} 
  \label{T1}
\end{table}

\begin{table}
\centering
\caption{Number of neurons in the neural networks}
\begin{tabular}{ |l|c|c|c| }
 \hline
  \textbf{Layer name} & \textbf{PN} & \textbf{CN} & \textbf{AN}\\
  \hline
Input Layer & 2K+2 & K & 2K\\
 \hline
Linear Layer  & 1000 & 100 & 1000 \\
 \hline
 ReLU Layer & 1000 & 100 & 1000\\
 \hline
Output Layer & 2K+2 & K & 2K\\
 \hline
\end{tabular}
\end{table}
In this section, we examine the numerical evaluation of the presented RSMA-CoMP MA-enabled model. We compare the results with several benchmark schemes defined as follows:

\begin{itemize}
  \item \textbf{CoMP RSMA - MA (fmincon)}: This benchmark examines the same system model presented above but with the optimization problem solved using the predefined MATLAB solver fmincon \cite{9586734}.
\item \textbf{CoMP RSMA - FPA}: It tackles the CoMP RSMA scenario by utilizing the traditional FPA on the user's side.  
\item \textbf{CoMP SDMA - MA}: This scheme involves a CoMP transmission system empowered by SDMA access technology and the users are equipped with MAs.
\item \textbf{CoMP SDMA - FPA}: Similar to the previous benchmark except that users use traditional FPA.
\end{itemize}

Unless specified otherwise, the primary simulation parameters are listed in Table I and the properties of the NNs are illustrated in Table II. In addition, the PRM matrix, and the azimuth and elevation angles of the AoA and AoD for the receiving and transmit paths, respectively are obtained based on \cite{10623434}. The simulation results are derived from 150 independent Monte Carlo trials.

In Fig. \ref{convv}, we demonstrate the convergence of the presented meta-learning approach where the graph studies the number of outer iterations for a single epoch versus the achievable sum rate value. At the beginning, the reward value decreased to almost 2. Then, it is clear that with the increase in the number of iterations, the reward value keeps on increasing until reaching a stable value at around 13.8 bps/Hz after 180 iterations. This signifies the model's ability to effectively update the optimization variables until reaching a close to optimal result. 
\begin{figure}[!t]
    \centering    \includegraphics[width=1.1\columnwidth]{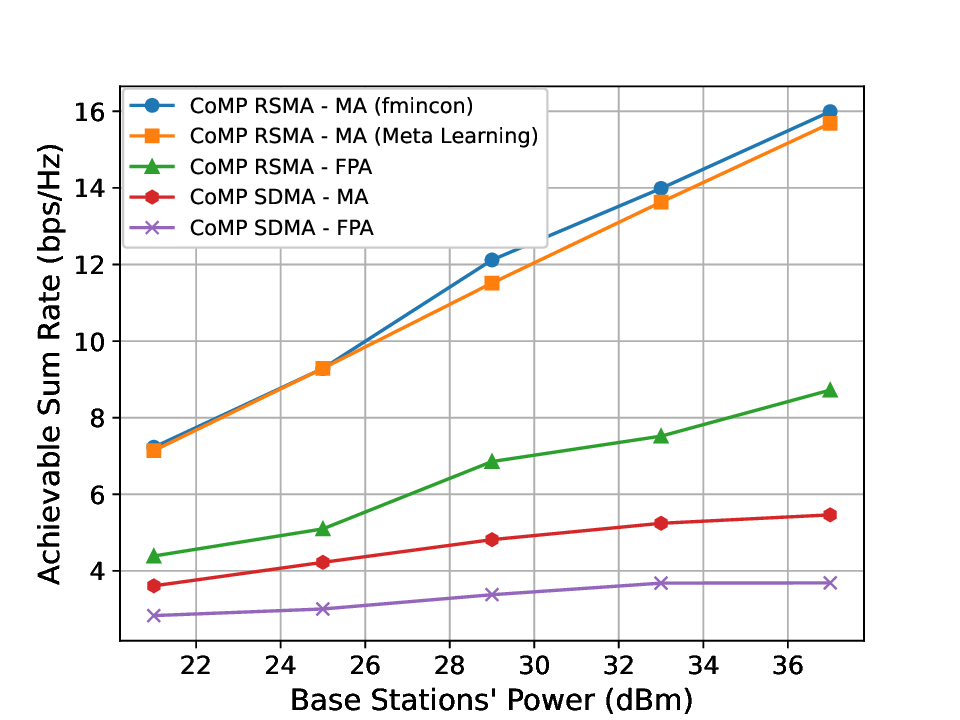}
    \caption{Achievable sum rate vs BSs power.}
	\label{bspower}
\end{figure}

\begin{figure}[!t]
    \centering    \includegraphics[width=1.1\columnwidth]{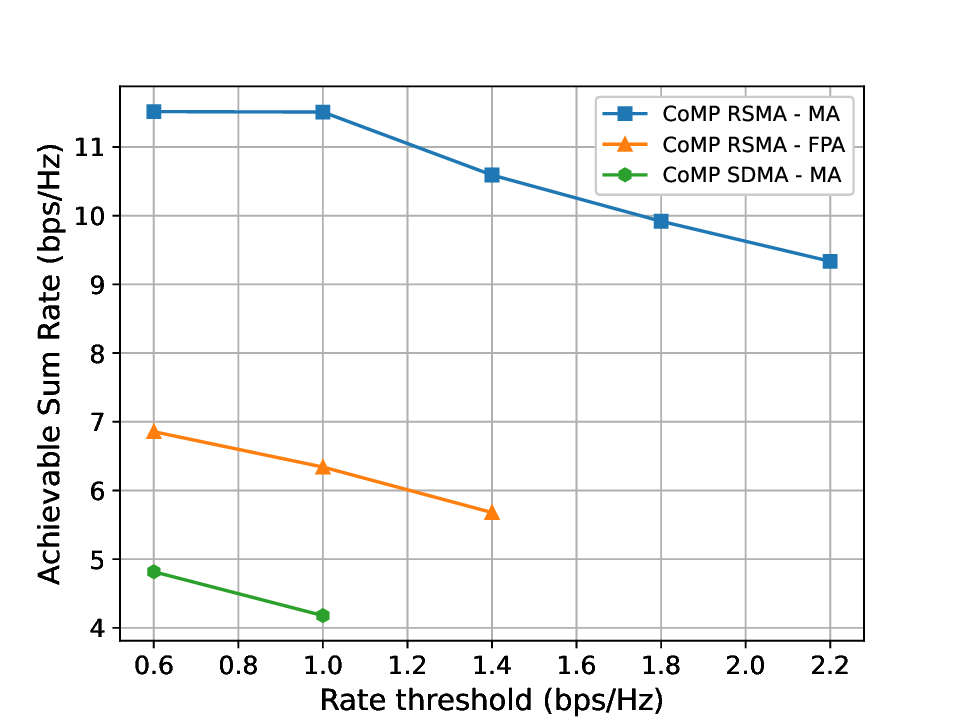}
    \caption{Achievable sum rate vs rate threshold  ($P_M^n = 29 $ $dBm,\forall n \in \mathcal{N} $).}
	\label{thresh1}
\end{figure}

Fig. \ref{bspower} examines the different BSs' power budget on the achievable sum rate of the presented model along with the other benchmarks. The graph presents an increase in the rate for all the lines with the increase in the power budget at the BSs with superiority for our CoMP-RSMA MA-enabled model, demonstrating the advantage of the MAs utilized at the users' side enabling effective mitigation of the fading effect through flexible antenna positioning. This is clear when comparing the presented scheme that supports MAs with a similar model utilizing traditional FPA, reaching gains up to 80\% when the power budget at the BSs is 37 dBm. Also, when comparing the CoMP-SDMA model that supports MAs with the one equipped with traditional FPA, achieving gains reaching 48\% at the same power budget. Thus, overcoming the problem of the upper limit due to the interference that constrains the rate even with high transmit power. 
In addition to the advantage of the MA technology, the adopted RSMA approach showed a major advantage over the SDMA technique. This is clear by comparing the RSMA and SDMA schemes that support MA technology, recording a gain reaching 180\%. The recorded gain can be interpreted by the ability to utilize the common stream, granting more freedom and better capability in managing the interference between the different users.

In addition to the previous advantages that were achieved due to the MA technology and RSMA technique, the proposed meta-learning algorithm demonstrated close to optimal results above 97 percent. This conclusion can be reached by comparing the results of the CoMP-RSMA model when solving the optimization problem with the meta-learning approach and when utilizing the fmincon solver using 100 initial points to guarantee obtaining very close to optimal results.

In Fig. \ref{thresh1}, the graph studies the rate threshold of the users versus the achievable sum rate for three different schemes. It is clear that with the increase in the rate requirements for the different users, the three schemes experience a drop in the rate. This is due to the need for the BSs to allocate more resources to some users, losing the advantage of increasing the rate for others. In addition, the presented model showed more robustness against the increase in the QoS requirements compared to the CoMP RSMA - FPA and CoMP SDMA - MA schemes that started yielding infeasible solutions at 1 and 1.4 bps/Hz, respectively. This is due to the ability of the MAs to battle fading and suppress its negative effect. Also, the RSMA technology grants advantages in terms of interference management and provides more degrees of freedom.

\begin{figure}[!t]
    \centering    \includegraphics[width=1.1\columnwidth]{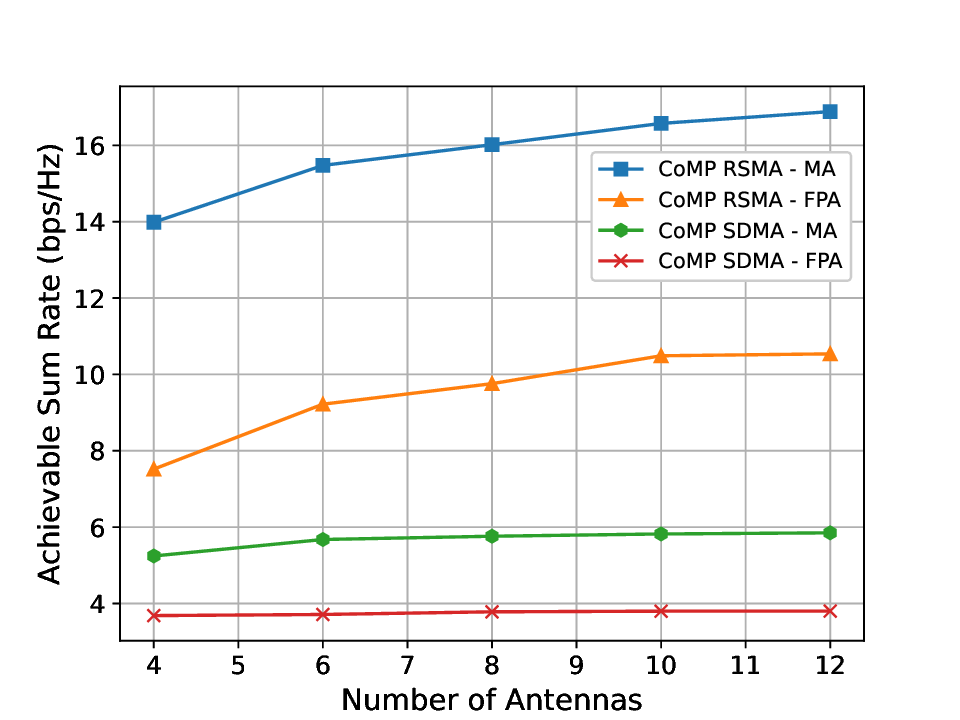}
    \caption{Achievable sum rate vs number of antennas.}
	\label{antennas}
\end{figure}

Fig. \ref{antennas} demonstrates the effect of the increase in the number of antennas at the BSs' side on the achievable sum rate at the users. 
The graph illustrates a significant improvement in the achievable sum rate for both RSMA schemes compared to a conservative increase in the two SDMA schemes (due to the interference caused by different users) using movable and fixed antenna systems. This can be explained by the additional degree of freedom granted by the additional antennas, allowing a better interference management process. In addition, we can observe that with the increase in the number of antennas at the BSs, the advantage of MAs decreases from about 87 \% at four antennas between the 2 RSMA schemes to about 60 \% at 12 antennas. We can interpret this decrease by the improved ability of the BSs to combat fading reducing the advantages provided by the MA on the users' side. 
\begin{figure}[!t]
    \centering    \includegraphics[width=1.1\columnwidth]{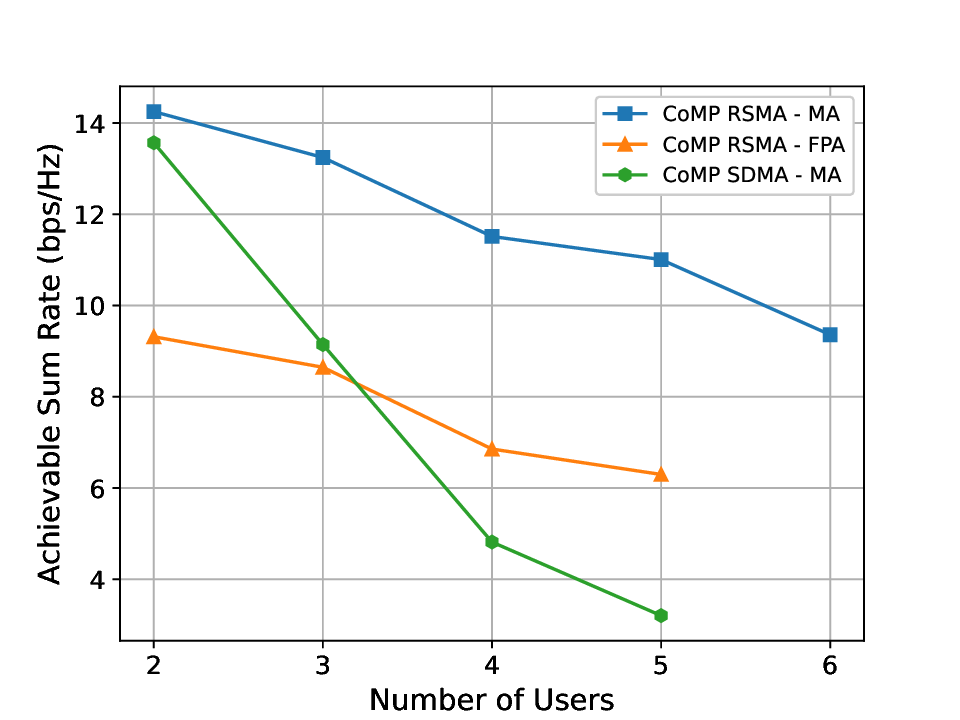}
    \caption{Achievable sum rate vs number of users  ($I = 6$).}
	\label{users}
\end{figure}

In Fig. \ref{users}, the graph presents the variation in the achievable sum rate as a function of the number of users in the system. All three schemes experienced a decrease in the achievable sum rate with the increase in the number of users. This can be explained by the need for the BS to split the power budget among more users, losing the advantage of increasing the sum rate of others. In addition, we can observe that the presented model is more robust to this increase, maintaining a feasible solution for six users compared to the other two benchmarks that failed to satisfy their requirements.

\begin{figure}[!t]
    \centering    \includegraphics[width=1.1\columnwidth]{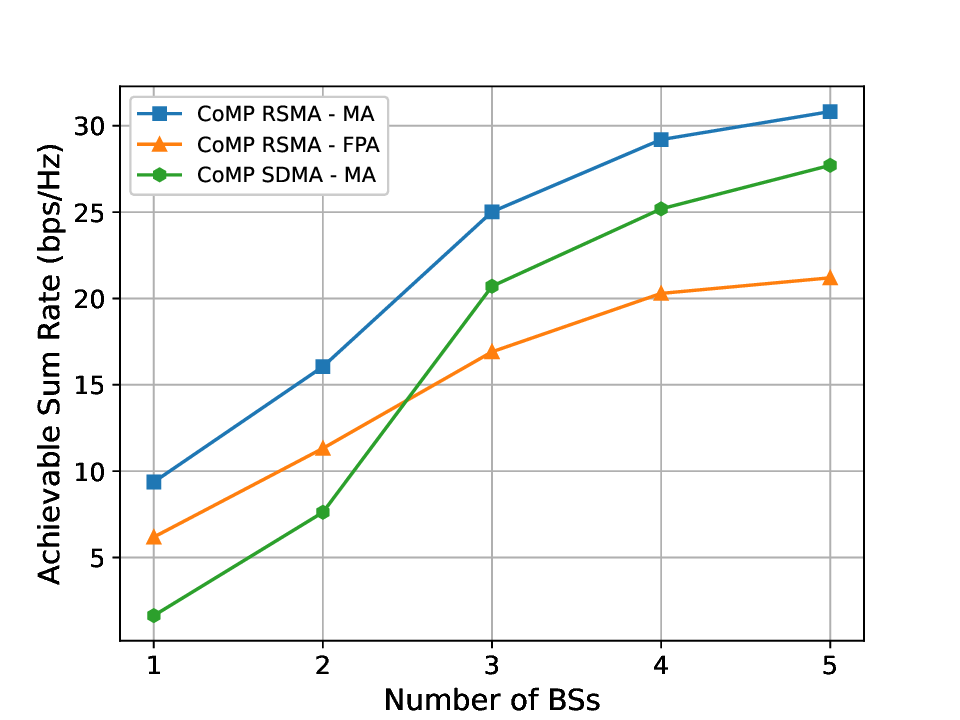}
    \caption{Achievable sum rate vs number of BSs($P_M^n = 37 dBm,\forall n \in \mathcal{N} $, $R_{th} = 0.4bps/Hs$).}
	\label{nBss}
\end{figure}

In Fig. \ref{nBss}, the achievable sum rate is studied versus the number of BSs for three different benchmarks. With the increase in the number of BSs, the achievable sum rate of the users of all the schemes increases with superiority for the presented model due to the additional transmission resources provided to the users. Moreover, following this increase, we can observe that the CoMP-SDMA model with MAs overcomes the CoMP-RSMA model with fixed antennas. This can be explained by the increased advantage of the MAs at the users in suppressing the fading compared to the benefit granted by the RSMA scheme. In addition, we can observe that the CoMP RSMA and SDMA schemes that support movable antennas reach close results at 5 BSs, signifying the additional degrees of freedom granted by the high number of BSs to the SDMA technique diminishing the advantage of RSMA technology. Finally, by comparing our model to the CoMP SDMA - MA when the achievable sum rate is 25 bps/Hz, we can realize that the proposed model is able to achieve the same rate with 3 BSs compared to 4 BSs for the other model. This highlights the advantage of the model in balancing between performance and complexity by achieving the same performance with fewer resources, an essential element for practical feasibility. 
\section{Conclusion}
In conclusion, in this paper, we studied a downlink CoMP-RSMA model in a scenario enabled by the MA technology on the users' side. MA technology offers great potential by optimizing the antennas' position, yielding better exploitation of the spatial diversity and enhanced ability to combat fading. In this regard, we formulate an optimization problem with the objective of maximizing the achievable sum rate of the communication users by determining the optimal beamforming vector at the BSs, the common stream portion for the users, and the antenna position at the users' side and respecting the quality of service constraints. This problem showed to be non-convex and intricate because of the high coupling of the different optimization variables. For that reason, we propose a gradient-based meta-learning algorithm that functions without prior training and can efficiently manage large-scale optimization problems. The numerical results demonstrated the advantage of the presented model in surpassing the upper-bound limitations on the achievable sum rate imposed by interference in the SDMA model, all while requiring fewer base stations, 
and the ability of the meta-learning solution to achieve close to optimal results (achieving over 97\% of the optimal performance).

\bibliographystyle{IEEEtran}
\bibliography{ref}

\end{document}